\documentclass[a4paper,12pt]{article}

\usepackage[english]{babel}
\usepackage[T1]{fontenc}
\usepackage{authblk}
\usepackage{natbib}
\usepackage{here}
\usepackage{setspace}
\usepackage{multirow}
\usepackage{lmodern}
\usepackage{amsmath, amssymb}
\usepackage{bm}
\usepackage{pdflscape}
\usepackage{authblk}
\usepackage{color}
\usepackage[table]{xcolor}
\usepackage{tikz}

\usepackage{booktabs}
\usepackage{afterpage}
\usepackage{textcomp}
\usepackage{float}
\usepackage{enumerate}
\usepackage{rotating}

\definecolor{babyblue}{rgb}{0.54, 0.81, 0.94}
\definecolor{babypink}{rgb}{0.96, 0.76, 0.76}

\usepackage[margin=1in]{geometry}

\usepackage{amsmath}
\usepackage{graphicx}
\usepackage[colorlinks=true, allcolors=blue]{hyperref} 

\usepackage{caption}
\usepackage{subcaption}

\definecolor{Gray}{gray}{0.85}
\newcolumntype{a}{>{\columncolor{Gray}}c}
\newcolumntype{d}{>{\columncolor{white}}c}

\begin{document}

    \begin{center}
        \vspace*{1cm}
        \large
	    \texttt{spAbundance}\textbf{: An R package for single-species and multi-species spatially-explicit abundance models}\\
         \normalsize
           \vspace{5mm}
	    Jeffrey W. Doser\textsuperscript{1, 2}, Andrew O. Finley\textsuperscript{2, 3, 4}, Marc K{\'e}ry\textsuperscript{5}, Elise F. Zipkin\textsuperscript{1, 2}
         \vspace{5mm}
    \end{center}
     \small
	      \textsuperscript{1}Department of Integrative Biology, Michigan State University, East Lansing, MI, USA \\
          \textsuperscript{2}Ecology, Evolution, and Behavior Program, Michigan State University, East Lansing, MI, USA \\
          \textsuperscript{3}Department of Forestry, Michigan State University, East Lansing, MI, USA \\
          \textsuperscript{4}Department of Statistics and Probability, Michigan State University, East Lansing, MI, USA \\
          \textsuperscript{5}Swiss Ornithological Institute, Sempach, Switzerland \\
          \noindent \textbf{Corresponding Author}: Jeffrey W. Doser, email: doserjef@msu.edu; ORCID ID: 0000-0002-8950-9895 \\
          \normalsize

\section*{Abstract}

\begin{enumerate}
    \item Numerous modeling techniques exist to estimate abundance of plant and animal populations. The most accurate methods account for multiple complexities found in ecological data, such as observational biases, spatial autocorrelation, and species correlations. There is, however, a lack of user-friendly and computationally efficient software to implement the various models, particularly for large data sets. 
    \item We developed the \texttt{spAbundance} \texttt{R} package for fitting spatially-explicit Bayesian single-species and multi-species hierarchical distance sampling models, N-mixture models, and generalized linear mixed models. The models within the package can account for spatial autocorrelation using Nearest Neighbor Gaussian Processes and accommodate species correlations in multi-species models using a latent factor approach, which enables model fitting for data sets with large numbers of sites and/or species.
    \item We provide three vignettes and three case studies that highlight \texttt{spAbundance} functionality. We used spatially-explicit multi-species distance sampling models to estimate density of 16 bird species in Florida, USA, an N-mixture model to estimate black-throated blue warbler (\textit{Setophaga caerulescens}) abundance in New Hampshire, USA, and a spatial linear mixed model to estimate forest aboveground biomass across the continental USA. 
    \item \texttt{spAbundance} provides a user-friendly, formula-based interface to fit a variety of univariate and multivariate spatially-explicit abundance models. The package serves as a useful tool for ecologists and conservation practitioners to generate improved inference and predictions on the spatial drivers of abundance in populations and communities.
\end{enumerate}

\textbf{Keywords}: Bayesian, imperfect detection, distance sampling, N-mixture model, hierarchical model

\newpage

\section{Introduction}

Understanding how abundance of plant and animal populations varies across space and time is a central objective in ecology and conservation management. A variety of sampling and associated modeling techniques have been developed over the last 50 years to estimate abundance while accounting for imperfect detection (i.e., the failure to observe all individuals of a species that are present at a location during the sampling period), including distance sampling and repeated counts, among others \citep{nichols2009inferences}. In distance sampling, the probability of detecting an individual is assumed to decay with increasing distance to the observer, which allows for the explicit estimation of abundance/density while accommodating imperfect detection of individuals \citep{buckland2001introduction}. Hierarchical distance sampling (HDS; \citealt{royle2004modeling}) extends classical distance sampling to model abundance/density as a function of spatially-varying covariates. \cite{royle2004modeling} introduced N-mixture models, which allow for estimation of abundance and effects of spatially-varying covariates while accounting for detection probability using replicated count data during some period where the population is assumed to be closed, i.e., where no births/deaths or immigration/emigration occur. In addition to approaches that explicitly account for imperfect detection, generalized linear mixed models (GLMMs) that estimate relative abundance (i.e., ignoring imperfect detection) can be used to assess how environmental covariates influence relative changes in abundance across space and/or time \citep{barker2018reliability, goldstein2022comparing}. Multi-species (i.e., multivariate) extensions of HDS \citep{sollmann2016hierarchical}, N-mixture models \citep{yamaura2012biodiversity}, and GLMMs (e.g., \citealt{hui2015model}) use count data from multiple species to estimate species-specific patterns in abundance, which may also estimate correlations between species in a joint species distribution model (JSDM) framework \citep{warton2015so}.

When modeling abundance across large spatial domains and/or using a large number of observed locations, accommodating spatial autocorrelation becomes increasingly important \citep{guelat2018}. Spatial autocorrelation can arise from a variety of ecological and/or biological processes, such as additional environmental drivers not included as covariates in the model, dispersal, species interactions, and source-sink meta-population dynamics (Chapter 9; \citealt{keryRoyle2021}). Failing to account for residual spatial autocorrelation (i.e., remaining spatial autocorrelation after accounting for environmental covariates) can lead to overly precise estimates and inferior predictive performance. Modeling spatial dependence is commonly done via the addition of spatially structured random effects to point-referenced spatial regression models (i.e., spatially-explicit models). Gaussian process-based random effects provide a flexible non-parametric approach to capture spatial patterns, offer unparalleled process parameter and predictive inference, and yield probabilistic uncertainty quantification. The hierarchical Bayesian framework is the preferred inferential framework for models developed here and in the literature due to their increased flexibility to fit models that would be infeasible with classical methods \citep{banerjee2014hierarchical}. Such models are, however, notoriously computationally intensive \citep{banerjee2012}, as computational complexity increases in cubic order with the number of spatial locations. These computational bottlenecks make fitting spatially-explicit models impractical for even moderately large data sets using Bayesian software packages such as \texttt{Stan} \citep{carpenter2017} and \texttt{NIMBLE} \citep{deValpine2017}.

Many popular, formula-based \texttt{R} packages exist that can fit various combinations of distance sampling models, N-mixture models, and/or spatially-explicit GLMMs for assessing spatial patterns in abundance (Supplemental Information S1: Table S1). The R package \texttt{unmarked} \citep{fiske2011, kellner2023unmarked} is commonly used to fit single-species distance sampling and N-mixture models, but cannot accommodate spatial autocorrelation. The \texttt{dsm} package \citep{miller2013spatial} can fit spatially-explicit distance sampling models using generalized additive models, the \texttt{hSDM} package \citep{hSDM} can fit spatially-explicit N-mixture models with an intrinsic conditional autoregessive model \citep{ver2018spatial}, while the \texttt{ubms} package \citep{kellner2021ubms} fits both spatially-explicit distance sampling and N-mixture models using restricted spatial regression \citep{hodges2010adding}. These packages, however, cannot accommodate multiple species within a multivariate framework. A variety of R packages exist to fit spatially-explicit univariate and multivariate GLMMs, such as \texttt{spBayes} \citep{finley2015spbayes}, \texttt{Hmsc} \citep{tikhonov2020joint}, and \texttt{sdmTMB} \citep{anderson2022sdmtmb}. However, none of these packages can explicitly account for imperfect detection. 

In this paper, we introduce the \texttt{spAbundance} R package for fitting Bayesian single-species and multi-species HDS models, N-mixture models, and GLMMs that may or may not include spatial autocorrelation in large data sets. We fit all spatially-explicit models with Nearest Neighbor Gaussian Processes (NNGPs), a computationally efficient approach that closely approximates a full Gaussian process while drastically reducing computational run times \citep{datta2016hierarchical, finley2019NNGP}. We designed \texttt{spAbundance} syntax to closely follow the syntax of \texttt{spOccupancy} \citep{doser2022spoccupancy}, an R package that fits a variety of analogous spatially-explicit occupancy models, which together provide a user-friendly and computationally efficient set of tools to model occupancy and abundance while accounting for spatial autocorrelation and imperfect detection.

\section{Overview of models in \texttt{spAbundance}}\label{models}

Next we give a brief overview of the models included in \texttt{spAbundance}. See Supplemental Information S1 for details on all prior distributions and their default values. 

\subsection{Single-species hierarchical distance sampling models}

The \texttt{spAbundance} functions \texttt{DS} and \texttt{spDS} fit non-spatial and spatially explicit single-species HDS models, respectively. Let $N(\bm{s}_j)$ denote the true abundance of a species of interest at site $j = 1, \dots, J$ with spatial coordinates $\bm{s}_j$. We model $N(\bm{s}_j)$ using either a Poisson or negative binomial (NB) distribution following
\begin{equation}\label{N-DS}
  \begin{split}
  N(\bm{s}_j) &\sim \text{Poisson}(\mu(\bm{s}_j)A(\bm{s}_j)) \text{, or, } \\
  N(\bm{s}_j) &\sim \text{NB}(\mu(\bm{s}_j)A(\bm{s}_j), \kappa),
  \end{split}
\end{equation}

where $\mu(\bm{s}_j)$ is the average abundance at site $j$, $A(\bm{s}_j)$ is an offset, and $\kappa$ is a positive dispersion parameter. Smaller values of $\kappa$ indicate overdispersion in the latent abundance values relative to a Poisson model, while higher values indicate minimal overdispersion in abundance. Note that as $\kappa \rightarrow \infty$, the negative binomial distribution becomes the Poisson distribution. The offset term $A(\bm{s}_j)$ can be used to convert $\mu(\bm{s}_j)$ to units of density (i.e., abundance per unit area), while if $A(\bm{s}_j) = 1$, $\mu(\bm{s}_j)$ is average abundance per site. We model $\mu(\bm{s}_j)$ using a log link function following 
\begin{equation}\label{mu-DS}
  \text{log}(\mu(\bm{s}_j)) = \bm{x}(\bm{s}_j)^\top\bm{\beta} + \text{w}(\bm{s}_j),
\end{equation}

where $\bm{\beta}$ is a vector of regression coefficients for a set of covariates $\bm{x}(\bm{s}_j)$ including an intercept, $\text{w}(\bm{s}_j)$ is a zero-mean spatial random effect, and the $\top$ denotes transposition of column vector $\bm{x}(\bm{s}_j)$. For non-spatial HDS models, $\text{w}(\bm{s}_j)$ is removed from Equation \ref{mu-DS}. For spatially-explicit HDS, we model $\textbf{w}(\bm{s})$ using a NNGP as a computationally efficient alternative to using a full spatial GP. More specifically, we assume that  
\begin{equation}\label{w}
  \textbf{w}(\bm{s}) \sim \text{Normal}(\bm{0}, \tilde{\bm{C}}(\bm{s}, \bm{s}', \bm{\theta}),
\end{equation}

where $\tilde{\bm{C}}(\bm{s}, \bm{s}', \bm{\theta})$ is a $J \times J$ NNGP-derived spatial covariance matrix and $\bm{\theta}$ is a vector of parameters governing the spatial process according to a spatial covariance function. \texttt{spAbundance} supports four spatial covariance models: exponential, spherical, Gaussian, and Mat\'ern \citep{banerjee2014hierarchical}. For the exponential, spherical, and Gaussian functions, $\bm{\theta} = \{\sigma^2, \phi\}$, where $\sigma^2$ is a spatial variance parameter controlling the magnitude of the spatial random effects and $\phi$ is a spatial decay parameter controlling the range of spatial autocorrelation, while the Mat\'ern function additionally includes a spatial smoothness parameter, $\nu$. See Supplemental Information S1 for statistical details on the NNGP approximation. 

Suppose observers count the number of individuals of the species of interest at each site $j$. Our HDS software implementation in \texttt{spAbundance} supports two types of ``sites'': line transects and point count surveys. In line transects, each site $j$ is a line transect the observer walks along and records the distance of each observed individual to the line within a set of $k = 1, \dots, K$ distance bands. In point count surveys, each site $j$ is the center of an imaginary circle at which an observer stands and records the distance of each observed individual to the center of the circle within $k = 1, \dots, K$ circular distance bands. Note that sometimes continuous distances are recorded rather than distance bins, in which case the continuous distance measurements can then be split into $K$ distance bins prior to analysis. Define $\bm{y}(\bm{s}_j)$ as a vector of $K$ values indicating the number of individuals observed within each distance band $k$ at site $j$. Similarly, let $\bm{y}^\ast(\bm{s}_j)$ be a vector of $K + 1$ values, where the first $K$ values correspond to $\bm{y}(\bm{s}_j)$, and the last value is the number of unobserved individuals at that location (i.e., $N(\bm{s}_j) - \sum_{k = 1}^Ky_k(\bm{s}_j)$). Note the last value in $\bm{y}^\ast(\bm{s}_j)$ is not observed (i.e., since $N(\bm{s}_j)$ is not known). We model $\bm{y}^\ast(\bm{s}_j)$ according to
\begin{equation}\label{yDS}
  \bm{y}^\ast(\bm{s}_j) \sim \text{Multinomial}(N(\bm{s}_j), \bm{\pi}_j^\ast),
\end{equation}

where $\bm{\pi}_j^\ast$ is a vector of cell probabilities with the first $K$ values denoted as $\bm{\pi}_j$ and the final value $\pi_{j, K + 1} = 1 - \sum_{k = 1}^K\pi_{j, k}$. More specifically, $\pi_{j, k}$ is the probability of detecting an individual in the $k$th distance band at site $j$. We define $\pi_{j, k}$ as 
\begin{equation}\label{piDS}
  \pi_{j, k} = \bar{p}_{j, k}\psi_{k},
\end{equation}

where $\bar{p}_{j, k}$ is the probability of detecting an individual in distance band $k$, given the individual occurs in distance band $k$, and $\psi_{k}$ is the probability an individual occurs in distance band $k$. The definitions of $\bar{p}_{j, k}$ and $\psi_k$ are different depending on whether the distance bands are linear (as in line transects) or circular (as in point count surveys). Following the standard distance sampling assumption that animals are uniformly distributed in space, for line transects we have 
\begin{equation}\label{psiLine}
   \psi_{k} = \frac{b_{k + 1} - b_k}{B}, 
\end{equation}

where $b_{k + 1}$ and $b_k$ are the upper and lower distance limits for band $k$, and $B$ is the line transect half-width (i.e., the maximum distance within which individuals are counted). Further, for distance $\text{x}$ we have
\begin{equation}\label{pBarLine}
  \bar{p}_{j, k} = \frac{1}{b_{k + 1} - b_k}\int_{b_k}^{b_{k+1}}g(\text{x})d\text{x}.
\end{equation}

For point count surveys, we have
\begin{equation}\label{psiPoint}
   \psi_k = \frac{b^2_{k + 1} - b^2_k}{B^2},
\end{equation} 

where $b_{k + 1}$ and $b_k$ are similarly the upper and lower distance limits for band $k$, and $B$ is the radius of the full point count circle. We then define $\bar{p}_{j, k}$ as 
\begin{equation}\label{pBarPoint}
  \bar{p}_{j, k} = \frac{1}{b^2_{k + 1} - b^2_k}\int_{b_k}^{b_{k+1}}g(\text{x})2\text{x}d\text{x}.
\end{equation}

For both line transects and point count surveys, $g(\text{x})$ is some declining function of distance $\text{x}$ from the transect line/point count survey center. We approximate the integrals in Equation \ref{pBarLine} and \ref{pBarPoint} using numerical integration. Our software implementation in \texttt{spAbundance} currently supports two detection functions: half-normal and negative exponential (see Supplemental Information S1 for their definitions). Both of these functions are governed by a scale parameter, $\sigma_j$, which can be modeled as a function of covariates to allow detection probability to vary across sites. More specifically, we have
\begin{equation}\label{sigmaDS}
  \text{log}(\sigma_j) = \bm{v}_j^\top\bm{\alpha},
\end{equation}

where $\bm{\alpha}$ is a vector of regression coefficients for covariates $\bm{v}_j$ (including an intercept). 

\subsection{Multi-species hierarchical distance sampling models}

Now consider the case where distance sampling data, $\bm{y}_i(\bm{s}_j)$, are collected for multiple species $i = 1, \dots, I$ at each survey location $j$ with coordinates $\bm{s}_j$. We are now interested in estimating the abundance of each species $i$ at each location $j$, denoted as $N_i(\bm{s}_j)$. We model $N_i(\bm{s}_j)$ analogous to Equation \ref{N-DS}, with expected abundance now varying by species and site according to
\begin{equation}\label{mu-msDS}
     \text{log}(\mu_i(\bm{s}_j)) = \bm{x}(\bm{s}_j)^\top\bm{\beta}_i + \text{w}^\ast_i(\bm{s}_j), 
\end{equation}

where $\bm{\beta}_i$ are the species-specific effects of covariates $\textbf{x}(\bm{s}_j)$ (including an intercept) and $\text{w}^\ast_i(\bm{s}_j)$ is a species-specific random effect. When $N_i(\bm{s}_j)$ is modeled using a negative binomial distribution, we estimate a separate dispersion parameter $\kappa_i$ for each species. We model $\bm{\beta}_i$ as random effects arising from a common, community-level normal distribution, which leads to increased precision of species-specific effects compared to single-species models \citep{sollmann2016hierarchical}. For example, the species-specific abundance intercept $\beta_{0, i}$ is modeled according to
\begin{equation}\label{betaComm}
     \beta_{0, i} \sim \text{Normal}(\mu_{\beta_0}, \tau^2_{\beta_0}),
\end{equation}

where $\mu_{\beta_0}$ is the community-level abundance intercept, and $\tau^2_{\beta_0}$ is the variance of the intercept across all $I$ species. The observation portion of the multi-species distance sampling model is identical to the single-species model and follows Equations \ref{yDS}-\ref{sigmaDS}, with all parameters indexed by species, and the species-specific coefficients $\bm{\alpha}_i$ modeled hierarchically analogous to the species-specific abundance coefficients $\bm{\beta}_i$ (Equation \ref{betaComm}). Given that species-specific effects are treated as random effects, such an approach requires at least 5-6 species in the modeled community in order for reliable estimation of variance parameters. 

\texttt{spAbundance} fits three types of multi-species models that differ in how they incorporate the species-specific random effect $\text{w}^\ast_i(\bm{s}_j)$ (if included). The function \texttt{msDS} fits the non-spatial multi-species distance sampling model of \cite{sollmann2016hierarchical} in which we remove the random effect $\text{w}_i^\ast(\bm{s}_j)$ from Equation \ref{mu-msDS}. The function \texttt{sfMsDS} fits spatial multi-species distance sampling models using a spatial factor model \citep{hogan2004bayesian}, which simultaneously accommodates spatial autocorrelation and residual species correlations in a spatial joint species distribution model framework. Briefly, we decompose $\text{w}^\ast_i(\bm{s}_j)$ into a linear combination of $q$ latent variables (i.e., factors) and their associated species-specific coefficients (i.e., factor loadings). More specifically, we have 
\begin{equation}\label{factorModel}
	\text{w}^\ast_i(\bm{s}_j) = \bm{\lambda}_i^\top\textbf{w}(\bm{s}_j),
\end{equation}

where $\bm{\lambda}_i^\top$ is the $i$\text{th} row of factor loadings from an $I \times q$ loadings matrix $\bm{\Lambda}$, and $\textbf{w}(\bm{s}_j)$ is a vector of length $q$ of independent spatial factors at site $j$. By setting $q \ll I$, we achieve dimension reduction to efficiently model communities with a large number of species \citep{taylor2019spatial, doser2023joint}. The approach accounts for residual species correlations via their species-specific responses to the $q$ spatial factors, which results in a residual interspecies covariance matrix that can be derived from the model as $\bm{\Sigma} = \bm{\Lambda}\bm{\Lambda}^\top$. We model each spatial factor using an independent NNGP according to Equation \ref{w}, except we fix the spatial variance parameter to 1 to ensure identifiability \citep{lopes2004bayesian}. As an alternative, the function \texttt{lfMsDS} models $\text{w}_i^\ast(\bm{s}_j)$ identical to Equation \ref{factorModel}, except assumes each of the $q$ factors in $\textbf{w}(\bm{s}_j)$ arises from an independent standard normal distribution. This model does not account for spatial autocorrelation but does allow for the estimation of species correlations. The models fit by \texttt{sfMsDS} and \texttt{lfMsDS} can be thought of as abundance-based JSDMs that account for imperfect detection (\citealt{tobler2019joint}; Chapter 8 in \citealt{keryRoyle2021}). 

Our factor modeling approach to fitting spatially-explicit multi-species models in \texttt{spAbundance} implicitly assumes species are correlated through latent factors $\textbf{w}(\bm{s}_j)$. If there is no interest in residual species correlations, we could imagine a multi-species model that includes a separate spatial process for each species. However, we do not include such models in \texttt{spAbundance} because they are computationally infeasible when working with even a moderate number of species (e.g., 10). Further, in the context of occupancy models, the spatial factor modeling approach has been shown to perform equally as well as a model that estimates a separate spatial random effect for each species even when there are no residual correlations between the species in the community \citep{doser2023joint}. 

\subsection{Single-species N-mixture models}

The functions \texttt{NMix} and \texttt{spNMix} fit non-spatial and spatial N-mixture models in \texttt{spAbundance}. Following the N-mixture model structure of \cite{royle2004n}, we assume observers count the number of individuals of a target species at each site $j$ over a set of multiple surveys $k = 1, \dots, K_j$, denoted as $y_k(\bm{s}_j)$. Note the number of surveys can vary by site, but at least some sites must be surveyed more than once to ensure identifiability. We model $y_k(\bm{s}_j)$ conditional on the true abundance of the species at site $j$, $N(\bm{s}_j)$, following 
\begin{equation}\label{N-NMix}
     y_k(\bm{s}_j) \sim \text{Binomial}(N(\bm{s}_j), p_{j, k}),
\end{equation}

where $p_{j, k}$ is the probability of detecting an individual given it is present at the site. We model $p_{j, k}$ using a logit link function in which we can allow detection probability to vary over space and/or surveys. More specifically, we have
\begin{equation} \label{p-NMix}
     \text{logit}(p_{j, k}) = \bm{v}_{j, k}^\top\bm{\alpha},
\end{equation}

where $\bm{\alpha}$ is a vector of effects of covariates $\bm{v}_{j, k}$ (including an intercept). The model for abundance $N(\bm{s}_j)$ is identical to the single-species distance sampling model, which can include covariates and/or spatial random effects (Equations \ref{N-DS}-\ref{w}).

\subsection{Multi-species N-mixture models}

Analogous to HDS models, we can extend single-species N-mixture models to model abundance of a community of $I$ total species \citep{yamaura2012biodiversity}. In multi-species N-mixture models, we estimate the abundance of species $i$ at spatial location $j$, $N_i(\bm{s}_j)$. Our model for $N_i(\bm{s}_j)$ follows that of multi-species HDS models, such that expected abundance can be modeled as a function of species-specific effects of spatially-varying covariates (Equation \ref{mu-msDS}) and species-specific random effects that can accommodate residual species correlations and residual spatial autocorrelation using a factor modeling approach (Equation \ref{factorModel}). Species-specific covariate effects are modeled hierarchically following Equation \ref{betaComm}. The observation portion of the multi-species N-mixture model is identical to the single-species model, now with species-specific covariate effects modeled hierarchically analogous to the abundance coefficients. \texttt{spAbundance} provides functions to fit non-spatial multi-species N-mixture models with (\texttt{lfMsNMix}) and without (\texttt{msNMix}) residual species correlations, as well as spatial multi-species N-mixture models that account for residual species correlations and spatial autocorrelation (\texttt{sfMsNMix}).

\subsection{Single-species GLMMs}

The functions \texttt{abund} and \texttt{spAbund} fit single-species (i.e., univariate) non-spatial and spatial GLMMs in \texttt{spAbundance} using abundance and related (e.g., biomass) data. As opposed to HDS and N-mixture models, GLMMs do not explicitly account for imperfect detection via an additional hierarchical component to the model, and instead directly model the observed abundance at site $j$, $y(\bm{s}_j)$ to provide inference on relative abundance (e.g., Chapter 1 in \citealt{keryRoyle2021}). Observed abundance $y(\bm{s}_j)$ is modeled using some probability distribution with mean $\mu(\bm{s}_j)$. \texttt{spAbundance} currently supports Poisson and negative binomial for use with count data and the Gaussian distribution for use with continuous abundance data, such as biomass. Mean relative abundance $\mu(\bm{s}_j)$ is modeled according to Equation \ref{mu-DS} for the Poisson and negative binomial cases, while the log link function is removed for the Gaussian case. Note that variables thought to influence detection probability can be incorporated in the model for $\mu(\bm{s}_j)$ to improve estimates of relative abundance (e.g., random observer effects, \citealt{link2002hierarchical}).

\subsection{Multi-species GLMMs}

Now consider the case where we have count data for multiple species $I$ at each survey location $j$, denoted $y_i(\bm{s}_j)$. We jointly model relative abundance of each species using a multivariate GLMM (e.g., \citealt{warton2015so, hui2015model}), in which expected abundance for each species $i$ at site $j$, $\mu_i(\bm{s}_j)$, is modeled analogous to Equations \ref{mu-msDS}-\ref{factorModel}. Note the log link function is removed from Equation \ref{mu-msDS} when modeling abundance using a Gaussian distribution. As with HDS and N-mixture models, \texttt{spAbundance} provides functions to fit non-spatial multivariate GLMMs with (\texttt{lfMsAbund}) and without (\texttt{msAbund}) residual species correlations. Multivariate spatial GLMMs with residual species correlations are fit using the \texttt{sfMsAbund} function.

\section{\texttt{spAbundance} functionality}

Here we highlight the five main tasks performed by \texttt{spAbundance} (see Table \ref{tab:coreFunctions} for function names). 

\textit{1. Data simulation.} The functions \texttt{simDS}, \texttt{simMsDS}, \texttt{simNMix}, \texttt{simMsNMix}, \texttt{simAbund}, and \texttt{simMsAbund} simulate data under the single-species and multi-species HDS, N-mixture, and GLMM frameworks for use in simulation studies or power analyses.

\textit{2. Model fitting.} Model fitting functions were described in Section \ref{models}. All models are implemented in a Bayesian framework using custom Markov chain Monte Carlo (MCMC) algorithms written in \texttt{C/C++} using \texttt{R}'s foreign language interface. \texttt{spAbundance} uses standard \texttt{R} formula syntax to specify abundance and detection probability models, with options to include random intercepts and random slopes using \texttt{lme4} syntax \citep{bates2015}. Users can specify initial values for the MCMC algorithm as well as each parameter's prior distribution to yield vague or informative priors as desired (Supplemental Information S1).

\textit{3. Model validation and comparison.} The function \texttt{ppcAbund} performs posterior predictive checks on \texttt{spAbundance} model objects to assess model Goodness of Fit. The function \texttt{waicAbund} computes the conditional version \citep{millar2018conditional} of the Widely Applicable Information Criterion (WAIC; \citealt{watanabe2010}) for model selection and assessment.

\textit{4. Posterior summaries.} We include \texttt{summary} functions for \texttt{spAbundance} model objects that display concise summaries of the posterior distributions for estimated parameters as well as the potential scale reduction factor ($\hat{\text{R}}$; \citealt{gelman1992inference}) and effective sample size for convergence diagnostics. Simple \texttt{plot} functions allow for further convergence diagnostics via visual assessment of traceplots. The complete posterior samples are returned as \texttt{coda::mcmc} objects \citep{coda} 

\textit{5. Prediction.} \texttt{predict} functions for all \texttt{spAbundance} model objects provide predictions of abundance across a user-specified set of locations, given covariate values and spatial coordinates. The resulting posterior predictive distributions can be used to generate abundance-based species distribution maps with associated uncertainty or to obtain population size estimates across the study region or within smaller areal units of interest. Users can also predict detection probability for HDS and N-mixture models to yield insight on how detection probability varies across a user-specified range of covariate values.

\section{Worked examples and online resources}

We demonstrate \texttt{spAbundance} functionality with three worked examples and three vignettes. Complete details for all worked examples are provided in Supplemental Information S1, along with associated code and data available on Zenodo \citep{dataAvail}. The vignettes are provided in Supplemental Information S2-S4 as well as on the package website (\url{https://www.jeffdoser.com/files/spabundance-web/}). Here we provide a short overview of the worked examples and vignettes.

\subsection{Case study 1: Bird density in central Florida}

\begin{figure}
    \centering
    \includegraphics[width=13cm]{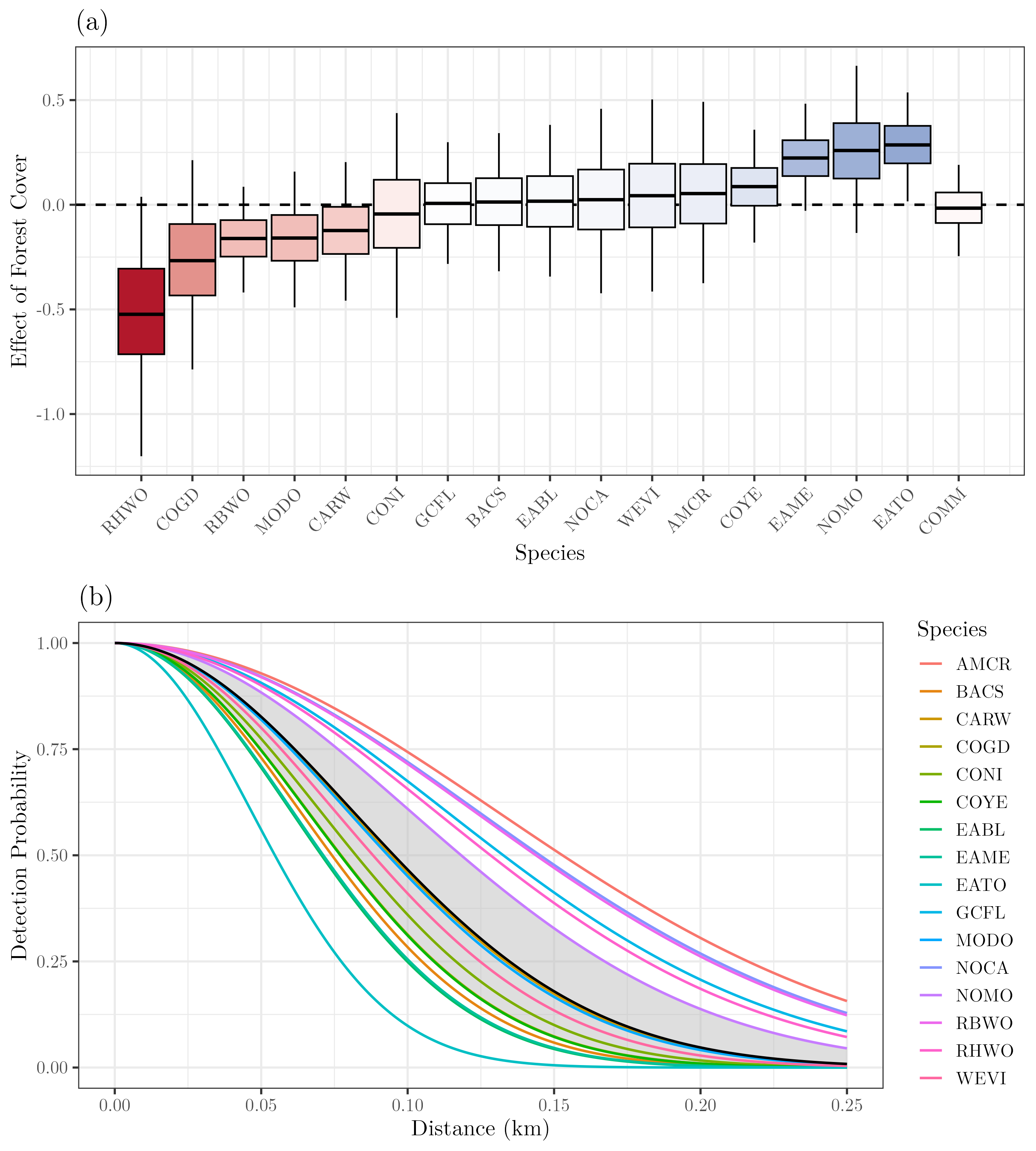}
    \caption{Species-specific effects of forest cover on density (a) and relationship between detection probability and distance from the observer (b) in the central Florida bird case study. Panel (a) shows the estimated mean (dark line), 50\% credible interval (box), and 95\% credible interval (whiskers) for the effect of forest cover on the overall community (COMM) and 16 individual species. In Panel (b), lines show the posterior mean detection probabilities for each species. The black line represents the average across the community (i.e., the community-level effect), and the grey region is the associated 95\% credible interval. See Supplemental Information S1 for species code definitions.}
    \label{fig:neonCovs}
\end{figure}

This case study demonstrates \texttt{spAbundance} functionality to fit spatial and nonspatial multi-species HDS models. We estimated density of 16 bird species in 2018 in the Disney Wilderness Preserve (48.5 km$^2$) in central Florida, USA. Distance sampling data were collected as part of the National Ecological Observatory Network landbird monitoring program \citep{neonData}. We compared the performance of the three multi-species model variants in \texttt{spAbundance} using WAIC. The spatial model substantially outperformed the non-spatial model with species correlations ($\Delta\text{WAIC} = 86$) and the non-spatial model without species correlations ($\Delta\text{WAIC} = 155$). Effects of forest cover on species-specific density varied across the community (Figure \ref{fig:neonCovs}a), resulting in clear spatial variation in density of the 16 species (Supplemental Information S1: Figure S1). Detection probability quickly decayed with increasing distance from the observer (Figure \ref{fig:neonCovs}b).  

\subsection{Case study 2: Black-throated blue warbler abundance in Hubbard Brook Experimental Forest}

In this case study, we showcase how to fit spatial and nonspatial single-species N-mixture models. We estimated abundance of black-throated blue warblers (\textit{Setophaga caerulescens}) in the Hubbard Brook Experimental Forest (31.8 km$^2$) in New Hampshire, USA using repeated count data from 2015 (\citealt{hbefData}; Supplemental Information S1). We found minimal support for overdispersion and residual spatial autocorrelation, with a non-spatial Poisson N-mixture model performing best according to WAIC among multiple candidate models. A strong negative quadratic relationship with elevation revealed that abundance peaked at mid-elevations in the forest (Supplemental Information S1: Figure S2). 

\subsection{Case study 3: Forest biomass across the continental USA}

\begin{figure}
    \centering
    \includegraphics[width=15cm]{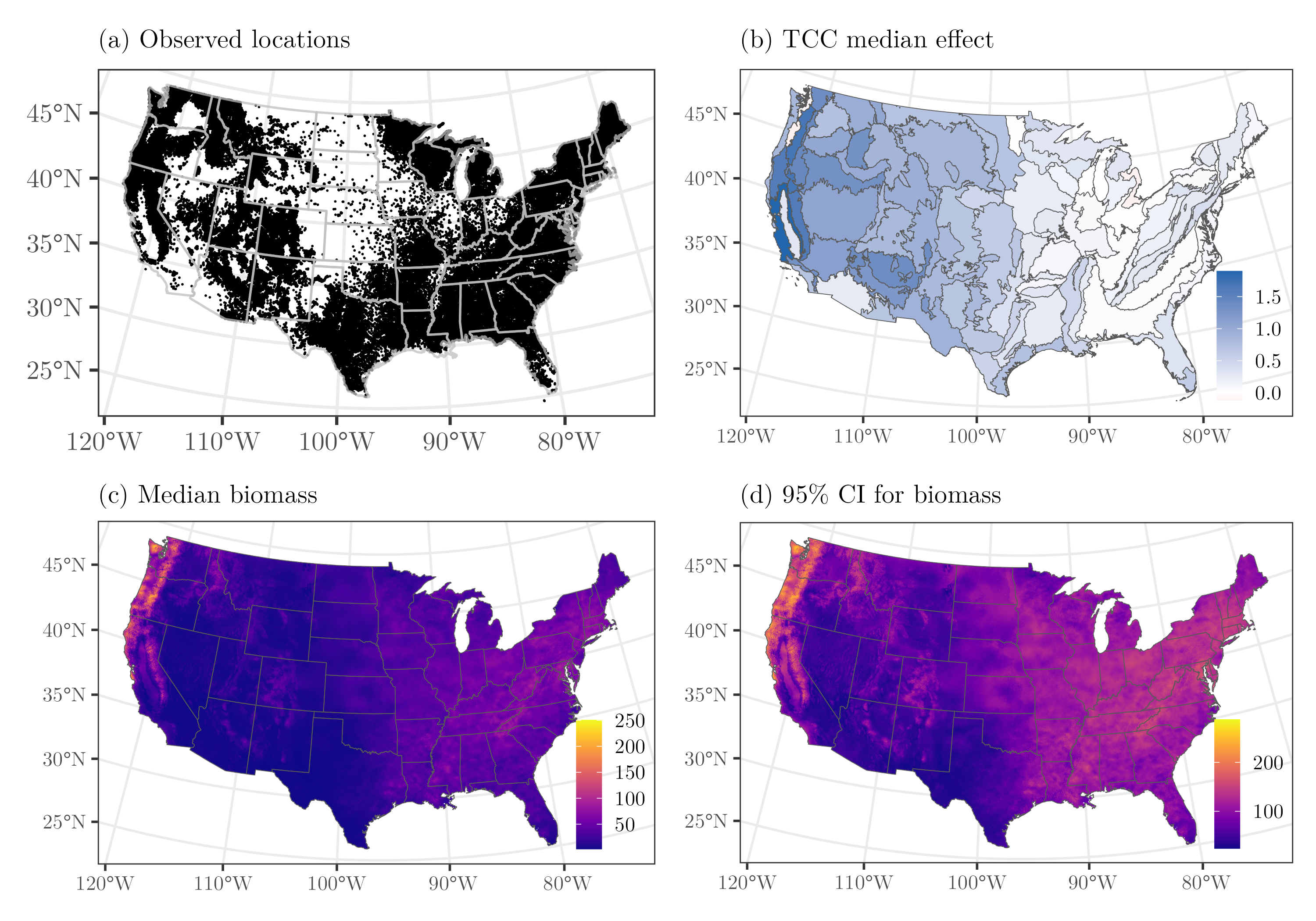}
     \caption{Data and predictions from the forest biomass case study. Panel (a) shows the observed locations of the 86,933 Forest Inventory and Analysis plots. Note these are the publicly available perturbed locations in which FIA adds a small amount of random noise to the true plot locations. Panel (b) shows the estimated random effect of tree canopy cover on forest biomass within distinct ecoregions. Panel (c) shows predicted biomass (posterior median) across the continental USA (tons per acre), with associated uncertainty (95\% credible interval [CI] width) depicted in Panel (d).}
    \label{fig:fiaResults}
\end{figure}

Our final case study demonstrates how \texttt{spAbundance} can be used to fit models using ``big data''. We estimated forest aboveground biomass across the continental US ($\sim 7.8$ million km$^2$) using data from $J = 86,933$ forest inventory plots (Figure \ref{fig:fiaResults}a) collected via the US Forest Service Forest Inventory and Analysis Program \citep{bechtold2005enhanced}. We fit a spatially-explicit univariate GLMM using a Gaussian distribution with an ecoregion-specific random slope of tree canopy cover to reflect potential spatial variation in the relationship between canopy cover and biomass across different forest types. We found an overall positive relationship between tree canopy cover and biomass (median = 0.54, 95\% credible interval 0.43-0.66), but clear variation in the magnitude of the effect across ecoregions (Figure \ref{fig:fiaResults}b). Biomass predictions across the US aligned with expectations, with highest biomass predicted in the Pacific Northwest (Figure \ref{fig:fiaResults}c,d).

\subsection{Vignettes}

The three package vignettes provide complete details and examples for fitting all single-species and multi-species model types for HDS models (Supplemental Information S2), N-mixture models (Supplemental Information S3), and GLMMs (Supplemental Information S4). We provide extensive details on the required data formats for implementing the models in \texttt{spAbundance} and all function arguments including their default values. We additionally provide code to manipulate resulting objects after fitting models to generate a variety of plot types and summary figures. 

\section{Conclusions and future directions}

We envision numerous extensions to existing \texttt{spAbundance} functionality and associated statistical methodology. We are currently working on functionality for zero-inflated models and spatiotemporal models, including ``generalized'' HDS and N-mixture models that account for imperfect availability \citep{chandler2011inference}. We encourage future simulation studies to better identify the potential impacts of spatial confounding on inference in spatially-explicit abundance models. Spatial confounding occurs when the spatial random effect is correlated with covariates included in the model, thus leading to difficulties in making inference on the covariate effects \citep{hodges2010adding}. While approaches exist to reduce such confounding (e.g., restricted spatial regression; \citealt{hodges2010adding}), they do not always provide more accurate inferences than standard approaches \citep{zimmerman2022deconfounding}. We echo the guidelines of \cite{makinen2022spatial} to assess the potential impacts of spatial confounding given the specific characteristics of the covariates of interest. 

The aim in developing \texttt{spAbundance} is to provide ecologists and conservation practitioners with a user-friendly tool to quantify and understand spatial variation in the abundance of plant and animal species. This \texttt{R} package fits Bayesian spatially-explicit single-species and multispecies versions of three of the most common modeling frameworks for ``unmarked'' data types: hierarchical distance sampling models, N-mixture models, and generalized linear mixed models. By using efficient statistical algorithms implemented in \texttt{C/C++} via R's foreign language interface, \texttt{spAbundance} is capable of handling data sets with a large number of species (e.g., >100) and locations (e.g., 100,000). Together, the package vignettes (Supplemental Information S2-S4), code to implement the three case studies \citep{dataAvail}, and the package website (\url{https://www.jeffdoser.com/files/spabundance-web/}) provide full details and thorough exposition of \texttt{spAbudance} model objects. 

\section{Author Contributions}

Jeffrey W. Doser developed the package with insights from Andrew O. Finley; Jeffrey W. Doser wrote the package vignettes with insights from Marc K\'ery; Jeffrey W. Doser performed analyses and led writing of the manuscript with critical insights from Elise F. Zipkin, Marc K\'ery, and Andrew O. Finley. All authors gave final approval for publication.

\section{Acknowledgements}
 
We thank J. Andrew Royle for helpful comments on the hierarchical distance sampling functionality and vignette. This work was supported by: Elise F. Zipkin NSF grants DBI-1954406 and DEB-2213565; Andrew O. Finley NASA CMS grants Hayes (CMS 2020) and Cook (CMS 2018), NSF grant DMS-1916395, joint venture agreements with the USDA Forest Service Forest Inventory and Analysis, USDA Forest Service Region 9 Forest Health Protection Northern Research Station. 

\section{Conflict of Interest}

We declare no conflict of interest.

\section{Data availability statement}

The package \texttt{spAbundance} is available on the Comprehensive R Archive Network (CRAN; \url{https://cran.r-project.org/web/packages/spAbundance/index.html}). Data and code used in the examples are available on GitHub (\url{https://github.com/zipkinlab/Doser_et_al_2024_MEE}) and Zenodo (\url{https://doi.org/10.5281/zenodo.10841651}; Doser et al. 2024).

\clearpage

\newpage

\section*{Tables and Figures}

\begin{table}[ht] 
  \begin{center}
    \caption{List of core functions in the \texttt{spAbundance} package. Model fitting function name components correspond to: \texttt{DS} (hierarchical distance sampling), \texttt{NMix} (N-mixture model), \texttt{abund} (abundance-based GLMM), \texttt{sp} (spatial), \texttt{ms} (multi-species), \texttt{lf} (latent factor), and \texttt{sf} (spatial factor).}
    \label{tab:coreFunctions}
    \begin{tabular}{ l  l }
      \hline
      Functionality & Description \\
      \arrayrulecolor{gray}\hline
      \textbf{Data simulation} & \\
      \arrayrulecolor{gray}\hline
      \texttt{simDS} & Simulate single-species distance sampling data \\
      \texttt{simMsDS} & Simulate multi-species distance sampling data \\
      \texttt{simNMix} & Simulate single-species repeated count data with imperfect \\ & detection \\
      \texttt{simMsNMix} & Simulate multi-species repeated count data with imperfect \\ & detection \\
      \texttt{simAbund} & Simulate single-species count data with perfect detection. \\
      \texttt{simMsAbund} & Simulate multi-species count data with perfect detection. \\
      \arrayrulecolor{gray}\hline
      \textbf{Model fitting} & \\
      \arrayrulecolor{gray}\hline
      \texttt{DS} & Single-species HDS model \\
      \texttt{spDS} & Single-species spatial HDS model \\
      \texttt{msDS} & Multi-species HDS model \\
      \texttt{lfMsDS} & Multi-species HDS model with species correlations \\
      \texttt{sfMsDS} & Multi-species spatial HDS model with species correlations \\
      \texttt{NMix} & Single-species N-mixture model \\
      \texttt{spNMix} & Single-species spatial N-mixture model \\
      \texttt{msNMix} & Multi-species N-mixture model \\
      \texttt{lfMsNMix} & Multi-species N-mixture model with species correlations \\
      \texttt{sfMsNMix} & Spatial multi-species N-mixture model with species correlations \\
      \texttt{abund} & Single-species GLMM \\
      \texttt{spAbund} & Single-species spatial GLMM \\
      \texttt{msAbund} & Multi-species GLMM \\
      \texttt{lfMsAbund} & Multi-species GLMM with species correlations \\
      \texttt{sfMsAbund} & Multi-species spatial GLMM with species correlations \\
      \arrayrulecolor{gray}\hline
      \textbf{Model assessment} & \\
      \arrayrulecolor{gray}\hline
      \texttt{ppcAbund} & Posterior predictive check using Bayesian p-values \\
      \texttt{waicAbund} & Compute Widely Applicable Information Criterion \\
    \hline
    \end{tabular}
  \end{center}	 
\end{table}

\newpage

\end{document}